\documentclass[a4paper,12pt]{article}
\usepackage{jheppub_modified_green}

\usepackage{placeins}

\usepackage{epsfig}
\usepackage[makeroom]{cancel}
\usepackage{amsfonts}
\usepackage{amssymb,esint}
\usepackage{enumitem}
\usepackage{comment}
\usepackage[normalem]{ulem}
\usepackage{amsthm}
\usepackage{subfig}
\usepackage{bm}
\usepackage{hyperref}
\usepackage{mathrsfs}
\usepackage{bbm}
\usepackage{cancel}
\usepackage{xcolor}
\usepackage{relsize}
\usepackage{float, slashed, graphicx, amssymb, amsmath}
\usepackage{shuffle}
\usepackage{pgfplots}
\usepackage{tikz-feynman}
\usepackage{silence}
\pgfplotsset{compat=1.18}
\tikzfeynmanset{compat=1.1.0}
\tikzfeynmanset{warn luatex=false}
\tikzfeynmanset{graviton/.style={circle, draw=green!60, fill=green!5, very thick, minimum size=7mm}}
\tikzfeynmanset{codot/.style={/tikz/shape=circle,/tikz/fill=white,/tikz/minimum size=0.1cm,/tikz/inner sep=1.8pt}}
\tikzfeynmanset{myblob/.style={/tikz/shape=rectangle,/tikz/fill=red,/tikz/minimum size=0.2cm,/tikz/inner sep=1.8pt} }
\tikzfeynmanset{ghc/.style={/tikz/shape=circle,/tikz/fill=white,/tikz/minimum size=0.05cm,} }
\tikzfeynmanset{HV/.style=
{/tikz/shape=circle,
/tikz/fill={rgb:black,1;white,2},
 /tikz/minimum size=0.01cm,/tikz/inner sep=3.0pt
 } }
\tikzfeynmanset{GR/.style={/tikz/shape=ellipse,/tikz/fill={rgb:black,1;white,2},/tikz/minimum size=0.3cm,} }
\tikzfeynmanset{GR2/.style={/tikz/shape=ellipse,/tikz/fill={rgb:black,1;white,2},/tikz/minimum width = 1.2cm, 
    /tikz/minimum height = 3.4cm} }
\tikzset{box/.pic={\filldraw[fill=black]  (0,0) circle (2.5pt); \filldraw [fill=black] (0.5,0) circle (2.5pt); \draw [line width=5pt] (0,0) -- (0.5,0);}}
 \tikzfeynmanset{myblob2/.style=
{/tikz/shape=rectangle,
/tikz/fill=black,
 /tikz/minimum width=0.1cm,/tikz/inner sep=1.8pt
 } }
 \tikzfeynmanset{sb/.style=
{/tikz/shape=circle,
/tikz/fill=black,
 /tikz/minimum size=0.3cm,/tikz/inner sep=1.pt
 } }
 \tikzfeynmanset{myblob/.style=
{/tikz/shape=ellipse,
/tikz/fill=red,
 /tikz/minimum width=0.5cm,
 } }

\usetikzlibrary{arrows.meta} 
\usetikzlibrary{calc}
\usetikzlibrary{decorations.pathmorphing}
\usetikzlibrary{decorations.pathreplacing}
\usetikzlibrary{decorations.markings}
\tikzset{
   vector2/.style={decorate, decoration={snake, amplitude=1pt, segment length=6pt}, draw,double},
   vector/.style={decorate, decoration={snake, amplitude=1pt, segment length=6pt}, draw},
	provector/.style={decorate, decoration={snake,amplitude=2.5pt}, draw},
	antivector/.style={decorate, decoration={snake,amplitude=-2.5pt}, draw},
    fermion/.style={draw=black, postaction={decorate},
        decoration={markings,mark=at position .55 with {\arrow[draw=black]{>}}}},
    fermionbar/.style={draw=black, postaction={decorate},
        decoration={markings,mark=at position .55 with {\arrow[draw=black]{<}}}},
    fermionnoarrow/.style={draw=black},
    gluon/.style={decorate, draw=black,
        decoration={coil,amplitude=4pt, segment length=5pt}},
    scalar/.style={dashed,draw=black, postaction={decorate},
        decoration={markings,mark=at position .55 with {\arrow[draw=black]{>}}}},
    scalarbar/.style={dashed,draw=black, postaction={decorate},
        decoration={markings,mark=at position .55 with {\arrow[draw=black]{<}}}},
    scalarnoarrow/.style={dashed,draw=black},
    electron/.style={draw=black, postaction={decorate},
        decoration={markings,mark=at position .55 with {\arrow[draw=black]{>}}}},
	bigvector/.style={decorate, decoration={snake,amplitude=4pt}, draw},
}
\tikzset{cross/.style={cross out, draw, 
         minimum size=2*(#1-\pgflinewidth), 
         inner sep=0pt, outer sep=0pt}}

\tikzstyle{block} = [draw, rectangle, 
    minimum height=3em, minimum width=6em]

\usepackage[T1]{fontenc} 

\def\sc#1{\overline{#1}}
\makeatletter
\newcommand \UPlus {\mathop {\operator@font \uplus }\limits }
\makeatother
\makeatletter
\newcommand \Bigcup {\mathop {\operator@font \bigcup }\limits }
\makeatother
\def\LabelNote#1{}
\def\Label#1{\label{#1}%
\smash{\hbox to\phipt{\raise1ex\hbox{\tiny[#1]}\hss}}}

\definecolor{bananayellow}{rgb}{1.0, 0.88, 0.21}
\definecolor{amber}{rgb}{1.0, 0.75, 0.0}

\newcommand{\cM}{\mathcal{M}}

\newcommand{\vecq}{\boldsymbol{q}}
\newcommand{\vecr}{\boldsymbol{r}}
\newcommand{\veca}{\boldsymbol{a}}
\newcommand{\vecp}{\boldsymbol{p}}
\newcommand{\veceps}{\boldsymbol{\epsilon}}

\def\nn{\nonumber}

\def\spa#1.#2{\left\langle#1\,#2\right\rangle}
\def\spb#1.#2{\left[#1\,#2\right]}

\def\be{\begin{equation}}
\def\ee{\end{equation}}
\def\bea{\begin{eqnarray}}
\def\eea{\end{eqnarray}}  
\allowdisplaybreaks

\usepackage{amssymb,amsmath}
\usepackage{mathtools} 
\usepackage{cancel} 
\usepackage{graphicx} 

\usepackage{hyperref}
\hypersetup{
	colorlinks=true,
	linktoc=page,
	citecolor=americanrose,
	linkcolor=cadmiumgreen,
	urlcolor=blue} 
\definecolor{americanrose}{rgb}{1.0, 0.01, 0.24}
\definecolor{cadmiumgreen}{rgb}{0.0, 0.42, 0.24}

\usepackage[colorinlistoftodos]{todonotes}

\newcommand{\Cdot}{{\cdot}} 
\makeatletter
\newcommand*{\bigcdot}{}
\DeclareRobustCommand*{\bigcdot}{%
  \mathbin{\mathpalette\bigcdot@{}}%
}
\newcommand*{\bigcdot@scalefactor}{.6}
\newcommand*{\bigcdot@widthfactor}{1.25}
\newcommand*{\bigcdot@}[2]{%
  \sbox0{$#1\vcenter{}$}
  \sbox2{$#1\cdot\m@th$}%
  \hbox to \bigcdot@widthfactor\wd2{%
    \hfil
    \raise\ht0\hbox{%
      \scalebox{\bigcdot@scalefactor}{%
        \lower\ht0\hbox{$#1\bullet\m@th$}%
      }%
    }%
    \hfil
  }%
}
\makeatother
\def\nn{\nonumber}

\title{Spinning  quadrupoles in effective field theories of gravity 
}
\author{Andreas Brandhuber$\mbox{}^{a}$,}
\author{Graham R.~Brown$\mbox{}^{b}$,}
\author{Gabriele Travaglini$\mbox{}^{a}$}
\author{\\and Pablo Vives Matasan$\mbox{}^{a}$}
\affiliation{$\mbox{}^{a}$Centre for Theoretical Physics, Department of Physics and Astronomy, \\
Queen Mary University of London, Mile End Road, London E1 4NS, United Kingdom}
\affiliation{$\mbox{}^{b}$Higgs Centre for Theoretical Physics, School of Physics and Astronomy, \\
The University of Edinburgh, Edinburgh EH9 3JZ, Scotland, United Kingdom}
\emailAdd{a.brandhuber@qmul.ac.uk}
\emailAdd{graham.brown@ed.ac.uk}
\emailAdd{g.travaglini@qmul.ac.uk}
\emailAdd{p.vivesmatasan@qmul.ac.uk}
\begin{document}
\begin{flushright}
	QMUL-PH-24-29
\end{flushright}

\abstract{
We study the  effect of the two independent  parity-even cubic interactions 
$I_1 = {R^{\alpha \beta}}_{\mu \nu} {R^{\mu \nu}}_{\rho \sigma} {R^{\rho \sigma}}_{\alpha \beta}$ and $
G_3 = I_1 -2 {R^{\mu \nu \alpha}}_\beta {R^{\beta \gamma}}_{\nu \sigma} {R^\sigma}_{\mu \gamma \alpha}$ 
on the spectrum of gravitational waves emitted in the quasi-circular inspiral phase of the merger of two spinning objects. 
Focusing on the aligned spin configuration, we extract the corrections to Newton's potential at linear order in the perturbations, using the four-point amplitude of the massive spinning objects evaluated in  the Post-Minkowskian expansion.
We then derive  the modifications to the quadrupole moments at leading order in the cubic perturbations, using a five-point amplitude with  emission of a soft graviton. These modified moments, along with the corresponding potentials, are then employed to calculate the power emitted by gravitational waves during the inspiral phase. Using these results, we  determine the   changes to the waveforms, up to linear order in spin, in the Stationary Phase Approximation.
Finally, we  comment on the relation between cubic and tidal perturbations.

}

\vspace{-2.6cm}

\maketitle

\flushbottom
 \tableofcontents
\newpage 

\section{Introduction}

The  direct detection of gravitational waves and the  observations of binary black hole mergers by the LIGO/Virgo/KAGRA collaboration have opened a new observational window, which may challenge our current understanding of gravity. With expectations of enhanced experimental sensitivity in  future observatories such as LISA, there will be a growing need for highly accurate theoretical predictions based on general relativity.
At the same time, one should also entertain the possibility that the Einstein-Hilbert  theory is modified by higher-derivative interactions in an effective field theory description. 
In this paper we will  consider 
the simplest such deformations, namely those cubic in the Riemann tensor,%
\footnote{It is known from the work of \cite{Tseytlin:1986zz,Deser:1986xr,Tseytlin:1986ti,AccettulliHuber:2019jqo} that corrections that are quadratic in the Riemann tensor do not affect scattering amplitudes.}
\begin{align}
I_1 &\coloneq  {R^{\alpha \beta}}_{\mu \nu} {R^{\mu \nu}}_{\rho \sigma} {R^{\rho \sigma}}_{\alpha \beta}\ , \qquad \ \ 
I_2 \coloneq {R^{\mu \nu \alpha}}_\beta {R^{\beta \gamma}}_{\nu \sigma} {R^\sigma}_{\mu \gamma \alpha}\ .  
\end{align}
Instead of    $I_2$,  it is more convenient to work with   the particular combination
\begin{align}
    G_3 \coloneq I_1 - 2 I_2\  .  
\end{align}
The reason is that in pure gravity its four-dimensional graviton amplitudes vanish \cite{AccettulliHuber:2020dal},%
\footnote{This was shown  at the level of the three-and four-point graviton amplitudes in \cite{vanNieuwenhuizen:1976vb} and \cite{Broedel:2012rc}, respectively.}
and we also note that this interaction is  topological in  six dimensions \cite{vanNieuwenhuizen:1976vb}. 
It does however contribute to amplitudes with scalars and gravitons \cite{Brandhuber:2019qpg,Emond:2019crr}, or massive spinning particles and gravitons \cite{Brandhuber:2024bnz}. 
Furthermore, the $G_3$ interaction can be mapped to a tidal deformation using a field redefinition \cite{AccettulliHuber:2020dal}.
The effective action we consider in this paper is thus
\begin{equation}
\begin{split}
\label{action}
S = \int\!d^4x \, \sqrt{-g} \,  \left( -\frac{2}{\kappa^2}R   \,  + \, \beta_1 I_1 + \beta_2 G_3 \right)
\, .
\end{split}
\end{equation}
We will not consider the two parity-odd cubic couplings $\tilde{I}_1$ and $\tilde{G}_3$, obtained from $\tilde{I}_1$ and $\tilde{G}_3$
by replacing one Riemann tensor by its dual, as they give rise to non-planar orbits \cite{Brandhuber:2024bnz}, while in this work we will focus on quasi-circular orbits.

Cubic deformations of gravity have been explored through various approaches, including general relativity methods \cite{Bueno:2016xff,Hennigar:2017ego,Silva:2022srr}, as well as amplitude techniques applied to  binary systems without spin \cite{Brandhuber:2019qpg, Emond:2019crr, AccettulliHuber:2020oou}, systems with a spinning black hole that is significantly heavier than the other \cite{Burger:2019wkq}, and more recently systems with two spinning black holes with arbitrary masses \cite{Brandhuber:2024bnz,Brandhuber:2024qdn}.
Related investigations into modified theories of gravity can be found in \cite{Sennett:2019bpc,deRham:2019ctd,deRham:2020ejn,deRham:2021bll,CarrilloGonzalez:2022fwg,Melville:2024zjq,Silva:2024ffz,Falkowski:2024bgb,Battista:2021rlh}. In particular, in \cite{AccettulliHuber:2020dal} amplitude methods were used to study the effects of cubic interactions \eqref{action} and tidal deformations on the dynamics of non-spinning black holes on quasi-circular orbits in the Post-Newtonian (PN) expansion. 
However, we now want to include spin  since generic black hole encounters will involve Kerr black holes, and study the effects on bound orbits in the PN expansion, also complementing our earlier work on cubic interactions for hyperbolic orbits \cite{Brandhuber:2024bnz,Brandhuber:2024qdn}.

Specifically, our first goal is  to  compute two quantities: first, the corrections to Newton's potential arising from the cubic deformations in the presence of spin; and then, the corrections to Einstein's formula for the power flux. The modified potentials  can be easily extracted from a Fourier transform of the two-to-two scattering amplitude of the massive spinning objects already computed in \cite{Brandhuber:2024bnz}. For the power flux we need
to compute the modifications to Einstein's quadrupole moment, which we derive from  a five-point  amplitude of the massive objects with the emission of a soft graviton.%
\footnote{In principle one could extract higher moments from our five-point amplitude, but  in this paper we will only focus on quadrupole corrections. }
We obtain the modified potential and  quadrupole to leading   Post-Minkowskian (PM) order and  for convenience we will quote results up to 7PN. 
We then combine our results to obtain the corrections to the gravitational waveforms describing the quasi-circular, adiabatic inspiral of binary black holes 
in the Stationary Phase Approximation. In order to focus on  planar motion we  restrict our attention to the aligned spin configuration, where the spins of the two black holes are orthogonal to the scattering~plane. We will also extend to the case of spinning black holes the  equivalence of a particular linear combination of the leading tidal perturbations  to the $G_3$ deformation, found  in  \cite{AccettulliHuber:2020dal} for scalar objects.   Finally, we discuss  some curious coincidences between the leading PN corrections to the potentials and momentum kicks from leading tidal and $I_1$ perturbations. 

The rest of the paper is organised as follows. 
 In Section~\ref{sec:2}
we briefly review the results of \cite{Brandhuber:2024bnz} for the one-loop four-point amplitude of four spinning particles in the presence of the interactions $I_1$ and $G_3$, from which we  extract the corrections to Newton's potential. We also introduce the point-particle effective action describing the conservative and dissipative dynamics of two spinning objects, which we specialise to the case of aligned spins. 
Section~\ref{sec:5ptAmps} begins with  the calculation of the five-point amplitude describing the inelastic scattering of two spinning objects accompanied by the emission of a graviton. From the soft limit of this amplitude we derive the corrections to the quadrupole moment to linear order in the perturbations, which for convenience we present up to quadratic order in the spins. 
We also  comment on the relation between cubic and  tidal deformations with and without spin. 
 Finally, in 
 Section~\ref{sec:AmplitudesToWaveforms}
we derive the corresponding corrections to the power emitted by the gravitational waves, as well as the corrections to the waveforms in the Stationary Phase Approximation.

\section{From cubic EFTs of gravity to worldlines}
\label{sec:2}
\subsection{Review of \texorpdfstring{$R^3$}{R^3} amplitudes}
In a recent paper \cite{Brandhuber:2024bnz}, the contributions from the cubic deformations to the four-point one-loop amplitudes of two spinning particles (with masses $m_{1,2}$ and spin vectors $a_{1,2}$) were found to be%
\footnote{In the following we  omit the coefficients of the cubic interactions $\beta_1$ and $\beta_2$, which will be reintroduced in Section~\ref{sec:AmplitudesToWaveforms}. 
We also mention that we define $G \coloneq \kappa^2 /(32\pi)$.}
\begin{align}
\label{eq:a4I1class}
     \cM_{4,I_1}
     &=i\left(\frac{\kappa}{2}\right)^6\frac{3}{16}m_1^3m_2^2|q|^3 \cosh(a_1\Cdot q)\bigg[ (\sigma^2{-}1)\cosh(a_2\Cdot q)- i \sigma\frac{\sinh(a_2\Cdot q)}{a_2\Cdot q} \epsilon(u_1 u_2 a_2 q)
 \bigg]\nn\\
 & + (1{\leftrightarrow}2)\,,\nn\\
\end{align}
and 
\begin{align}
\begin{split}
\label{eq:a4G3class}
    \cM_{4,G_3}
    &=-i\left(\frac{\kappa}{2}\right)^6\frac{3}{16}m_1^3m_2^2 |q|^3 \cosh(a_1\Cdot q)\cosh(a_2\Cdot q)+ (1\leftrightarrow 2)\,.
\end{split}
\end{align}
\begin{figure}[h]
\begin{equation*}\label{eq: kinematics}
		\begin{tikzpicture}[scale=14,baseline={([yshift=-1mm]centro.base)}]
			\def\x{0}
			\def\y{0}

			\node at (0+\x,0+\y) (centro) {};
			\node at (-3pt+\x,-3pt+\y) (uno) {\footnotesize $p_1$};
			\node at (-3pt+\x,3pt+\y) (due) {\footnotesize $p_2$};
			\node at (3pt+\x,3pt+\y) (tre) {\footnotesize $p_2^\prime=p_2- q$};
			\node at (3pt+\x,-3pt+\y) (quattro) {\footnotesize $\ \ p_1^\prime = p_1+q$};

			\draw [thick,double,blue] (uno) -- (centro);
			\draw [thick,double,amber] (due) -- (centro);
			\draw [thick,double,amber] (tre) -- (centro);
			\draw [thick,double,blue] (quattro) -- (centro);

			\draw [->] (-2.8pt+\x,-2pt+\y) -- (-1.8pt+\x,-1pt+\y);
			\draw [<-] (2.8pt+\x,-2pt+\y) -- (1.8pt+\x,-1pt+\y);
			\draw [<-] (-1.8pt+\x,1pt+\y) -- (-2.8pt+\x,2pt+\y);
			\draw [->] (1.8pt+\x,1pt+\y) -- (2.8pt+\x,2pt+\y);

			\shade [shading=radial] (centro) circle (1.5pt);
		\end{tikzpicture}
\end{equation*}
\caption{Kinematics of the four-point one-loop amplitude, where  the bottom (top)  particles have mass $m_1$ ($m_2$) and classical spin vector $a_1$ ($a_2$). 
}
\label{fig:1loop}
\end{figure}
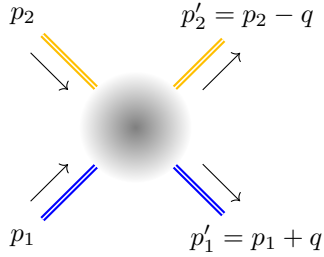
The kinematics is  shown in Figure~\ref{fig:1loop}, and we  have also introduced the ring radius vector $a^\mu \coloneq S^\mu / m$
\cite{Cangemi:2022abk}, where 
$S^\mu$ is the spin~\cite{Vines:2017hyw,Guevara:2018wpp,Chung:2019yfs,Bern:2020buy,Bautista:2021wfy}, with  
\begin{align}
    p\Cdot a(p) =0\, . 
\end{align}  
We have also introduced 
\begin{align}
\sigma\coloneq \frac{p_1 \Cdot p_2}{m_1 m_2}\, . 
\end{align}
Often we will work in the centre of mass frame, where we parameterise the momenta of the  particles as  
\begin{align}
\label{paraCOM}
\begin{split}
p_1 &= \left(E_1, \vecp -\frac{\vecq}{2}\right)    \, ,  
\qquad  
p_2 = \left(E_2, - \vecp +\frac{\vecq}{2}\right) \, , 
\\
p_1^\prime &= \left(E_1, \vecp +\frac{\vecq}{2}\right)    \, ,  
\qquad  
p_2^\prime  = \left(E_2, - \vecp -\frac{\vecq}{2}\right)
\, , 
\end{split}
\end{align}
with $E_1=\sqrt{m_1^2 + \vecp^2 + \vecq^2/4}$, $E_2=\sqrt{m_2^2 + \vecp^2 + \vecq^2/4}$, with $\vecp \Cdot \vecq=0$. Note that in the classical limit $E_i \to \sqrt{m_i^2+\vecp^2}$.

The amplitudes  in \eqref{eq:a4I1class} and \eqref{eq:a4G3class} were obtained by gluing   the linearised energy-momentum tensor of a  Kerr black hole \cite{Vines:2017hyw,Guevara:2018wpp,Chung:2019yfs,Bern:2020buy,Bautista:2021wfy}, 
\begin{equation}
\label{eq:Tmunu}
 T^{\mu \nu} = \left( \frac{\kappa}{2} \right) 2 \left[ \cosh (a\Cdot q) p^\mu p^\nu  - i \frac{\sinh (a\Cdot q)}{a \Cdot q} p^{(\mu} \epsilon ^{\nu)}_{\ \alpha \beta \gamma}p^\alpha a^\beta q^\gamma\right]    \, , 
\end{equation}
to the three-graviton vertices resulting from the $R^3$ interactions, which can be easily obtained by linearising the Riemann tensors and expanding in the metric fluctuations. However, for the purpose of this paper it will prove more useful to use these cubic vertices keeping one Riemann tensor, namely the one corresponding to the radiated, on-shell graviton, unexpanded. In the all-outgoing convention, that we adopt, the $I_1$ vertex with two off-shell legs takes the form
\begin{equation}
\label{eq:VertexI1} 
\begin{tikzpicture}[line width=1.5 pt,scale=.5,baseline={([yshift=-0.4ex]current bounding box.center)}]
	\draw[double, vector, line width = .75] (0:0) -- (0:2);
	\draw[double, vector, line width = .75] (0:0) -- (120:2);
	\draw[double, vector, line width = .75] (0:0) -- (240:2);
        \node at (60:1.2) {$I_1$};
        \node at (120:2.6) {$1,\alpha\beta$};
        \node at (240:2.6) {$2,\gamma\delta$};
\end{tikzpicture}
= 
48 i\, \left(\frac{\kappa}{2}\right)^2 k_1^\mu \left(k_1\Cdot k_2 \eta^{\alpha\gamma} - k_1^\gamma k_2^\alpha\right) k_2^\nu R_{\beta\mu\delta\nu}
\ ,
\end{equation}
with $R^{\mu\nu\rho\sigma}=(\kappa/2) (k^\mu \varepsilon^\nu-k^\nu \varepsilon^\mu)(k^\rho \varepsilon^\sigma-k^\sigma \varepsilon^\rho)$,  with  similar, but more complicated, expressions for $I_2$ and $G_3$.%
\footnote{The expression for the $I_2$ vertex is given in Appendix~\ref{sec:I2vertex}.}

\subsection{Quadrupoles and the PN formalism}

In the PN framework, one can describe the conservative and dissipative dynamics of two objects with  mass $m_1$ and $m_2$ coupled to
the gravity effective action \eqref{action},
using  the  point-particle effective action \cite{Goldberger:2004jt,Endlich:2017tqa}
\begin{eqnarray}
\label{correctedL}
S_{\text{pp}}  & = & \int\!dt \, \left[ \frac{1}{2} \mu \, \dot{\vecr}^{\, 2} \, - \, V(\vecr,\boldsymbol{p}) \, +\,  \frac{1}{2} Q^{ij}(\vecr, \boldsymbol{p}) R^{0i0j} + \cdots\right] \ . 
\end{eqnarray}
Here $\mu\coloneq m_1 m_2/ (m_1 + m_2)$
 denotes  the reduced mass,   and  $\vecr(t)$ is  the relative position of the two bodies. 
 The potential $V\big(\vecr, \boldsymbol{p}\big)$  to first order in the parameters $\beta_i$ will be computed in the next section; $Q^{ij}\big(\vecr, \boldsymbol{p}\big)$ is the quadrupole moment, which we will later extract from the classical five-point amplitude.   
 The dots in \eqref{correctedL} represent higher-order terms that we will not need  in the rest of the paper.  
We also note that \eqref{correctedL} can be trusted in the inspiral phase,  before the objects reach relativistic velocities.

In general, the addition of spin means that orbits become non-planar. However, when these spins are aligned perpendicular to the scattering plane, the motion remains in the plane%
\footnote{ As we have seen in \cite{Brandhuber:2024bnz}, the motion is non-planar in the presence of the parity-odd interactions $\tilde{I}_1$ and $\tilde{G}_3$, obtained from $I_1$ and $G_3$ by replacing one Riemann tensor by its dual.}. This is called the \textit{aligned spin} configuration,
\begin{equation}
    p_i \Cdot a_j = 0\, , \qquad {a}_i\Cdot r =0\, , \qquad 
i,j=1,2\, . 
\end{equation}
The scattering plane can, in general, be arbitrary but for concreteness we will consider scattering in the $xy-$plane, with the spins pointing along the $\hat{z}$ direction. With these conventions, we can specialise the  kinematics of the binary in the centre of mass frame given in \eqref{paraCOM} to
\begin{align}
    p_1 &= (E_1,-|\vecp| \sin\theta,|\vecp|\cos\theta,0)\ , \nn\\
    p_2 &= (E_2,|\vecp| \sin\theta,-|\vecp|\cos\theta,0)\ , \nn\\
    a_i &= (0,0,0,|\veca_i|)\ , \nn\\
    r &= (0,|\vecr|\cos\theta,|\vecr|\sin\theta,0)\ .
\end{align}
When it is clear from context, we will write $r \coloneq |\vecr|$ and $a_i \coloneq |\veca_i|$.

\subsection{Spinning cubic corrections to the potential}\label{sec:CorrectionsPotential}

The  potential in \eqref{correctedL} can by obtained from the   elastic one-loop amplitudes of \eqref{eq:a4I1class} and \eqref{eq:a4G3class}, by Fourier transforming to position  space with an appropriate  normalisation~\cite{Iwasaki:1971vb},
\begin{equation}\label{eqn:PotentialNormFactor}
    V(\vecr,\boldsymbol{p}) = \int\! \frac{d^3 \vecq}{(2\pi)^3}\ e^{-i \vecq\Cdot\vecr}\frac{i\cM_4(q)}{4 E_1 E_2}\ .
\end{equation}
We can recast the hyperbolic functions as shift operators by using the fact that, inside the integral, $\vecq\to i\nabla_{\vecr}$. This allows us to remove them from the integral and apply them as shifts later,
\begin{align}
    \int\!\frac{d^3 \vecq}{(2\pi)^3}\ \cosh{(\veca_i\Cdot \vecq)} e^{-i \vecq\Cdot\vecr} f(\vecq) &\to \frac{1}{2}\left(e^{i \veca_i\Cdot\nabla_{\vecr}}+e^{-i \veca_i\Cdot\nabla_{\vecr}}\right) \int\!\frac{d^3 \vecq}{(2\pi)^3}\ e^{-i \vecq\Cdot\vecr} f(\vecq) \nn \\
    &= \frac{1}{2}\left(\tilde{f}(\vecr+i\veca_i)+\tilde{f}(\vecr-i\veca_i)\right)\ , \label{eq:HyperbolicTrick}
\end{align}
and similarly for $\sinh(\veca_i\Cdot\vecq)$. We can deal with the Levi-Civita term by a similar shifting procedure, defining $\veceps_2\Cdot\vecq = \epsilon(u_1 u_2 a_2 \vecq)$:
\begin{align}
    \int\!\frac{d^3\vecq}{(2\pi)^3}\  e^{-i \vecq\Cdot\vecr}\veceps_2\Cdot\vecq f(\vecq) &= \left.\frac{d}{dt} \int\!\frac{d^3\vecq}{(2\pi)^3}\  e^{-i \vecq\Cdot(\vecr+it\veceps_2)} f(\vecq)\right|_{t=0} \nn \\
    &= \frac{d}{dt} \left.\tilde{f}(\vecr + it\veceps_2)\right|_{t=0} \, . \label{eqn:DerivativeTrick}
\end{align}
This leaves us with the computation of two integrals,
\begin{align}
    \label{eq:FourierIntegral1}
    &\int\! \frac{d^3\vecq}{(2\pi)^3} e^{-i\vecq\Cdot\vecr} |\vecq|^3\ , \\
    \label{eq:FourierIntegral2}
    &\int\! \frac{d^3\vecq}{(2\pi)^3} e^{-i\vecq\Cdot\vecr} |\vecq|^3\frac{\sinh{(\veca_2\Cdot \vecq)}}{\veca_2\Cdot\vecq}\ ,
\end{align}
where we have kept the $\sinh{(\veca_2\Cdot \vecq)}$ term to make the cancellation of the spurious pole manifest. These then have to be shifted and differentiated as appropriate. The first integral has a simple closed-form expression, and the second can be done by using the Schwinger trick as in \cite{Brandhuber:2024bnz}. The explicit expressions for the integrals are presented in Appendix~\ref{sec:FourierIntegrals}. 

Once the dust settles, the potentials for the various cubic corrections evaluated in the centre of mass frame, for the aligned spin configuration,  are given by 
\begin{align}
\label{eqn:I1SpinningPot}
    V_{I_1}(r,\vecp) &= -  \left(\frac{\kappa }{2}\right)^6\frac{9 }{\pi ^2 r^6}\Biggl[%
    \frac{\left(m_1+m_2\right)^3 \vecp^2}{16 m_1 m_2}+\frac{3 \left(m_1+m_2\right) \sqrt{\vecp^2} \left(a_2 m_1+a_1 m_2\right)}{8 r}\nn \\ & -
    \frac{\left(m_1-m_2\right)^2 \left(m_1+m_2\right)^3 \vecp^4}{32 m_1^3 m_2^3}
    \Biggr]+ \cdots \, , 
    \,  
    \\
    \label{eqn:G3SpinningPot}
    V_{G_3}(r,\vecp) &= \left(\frac{\kappa}{2}\right)^6 \frac{9 m_1 m_2 (m_1+m_2)}{16\pi^2 r^6} \Biggr[1-\vecp^2\frac{m_1^2+m_2^2}{2m_1^2 m_2^2} +
 \vecp^4\frac{3m_1^4 + 2 m_1^2 m_2^2 + 3 m_2^4}{8 m_1^4 m_2^4}  \nn \\ 
 &+ 
    \frac{3(a_1^2+a_2^2)}{r^2}\Biggr]+\cdots 
    \ , 
\end{align}
where $a_i \coloneq |\veca_i|$, $i=1,2$ and $r \coloneq |\vecr|$.
We have expanded these expressions in PN and spin for presentation, but the integrals in Appendix \ref{sec:FourierIntegrals} are all-order expressions.%
\footnote{For the sake of organising the PN expansion, we remind that because of the virial theorem, each power of $1/r$ corresponds to a power of $v^2$, that is one more PN order. We also  note that 
each power of the ring radius $a$ is always 
accompanied by a power of $1/r$.} For the sake of presentation we have quoted terms up to   7PN.   
The complete answer at next-to-leading order in PN and spin will involve contributions from higher-loop diagrams which we do not consider here.

Two comments are in order here. First, we note that 
the first corrections to the potential  from $I_1$ and $G_3$ appear at 6PN and 5PN order, respectively. We should however stress that these are only formally very high PN orders. The corrections are dressed with the couplings $\beta_{1,2}$ which lead to large numerical coefficients that effectively reduce the PN order, see also \cite{Sennett:2019bpc} for a related discussion in the context of quartic curvature corrections. 

Furthermore, 
as expected from the amplitudes, the $G_3$ 
 potential  
only contains corrections that are  even  in spin, while the $I_1$ 
potential 
contains the Levi-Civita term which contributes at odd powers too. It is this term which is also responsible for contributions to half PN orders, as
\begin{equation}
    \veceps_i\Cdot\vecr = -\frac{m_1+m_2}{m_1 m_2} a_i r \sqrt{\vecp^2} \, . 
\end{equation}
in the aligned spin configuration and  as $\vecp^2\rightarrow0$.
We can also see that each power of spin brings along with it one additional power of $1/r$, as in the amplitude each spin vector always comes accompanied by $q$, which is conjugate to $r$.

\section{Five-point amplitudes and spinning quadrupole moments}\label{sec:5ptAmps}
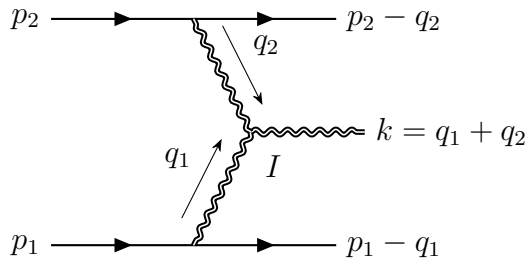
\begin{figure}
\centering
\begin{tikzpicture}[scale=1.5]
\begin{feynman}
    \vertex (3g) at (0.5,0);
    \vertex (p2g) at (0,1);
    \vertex (p1g) at (0,-1);
    \vertex (p1) at (-1.25,-1)[left]{$p_1$};
    \vertex (p1p) at (1.25,-1)[right]{$p_1-q_1$};
    \vertex (p2) at (-1.25,1)[left]{$p_2$};
    \vertex (p2p) at (1.25,1)[right]{$p_2-q_2$};
    \vertex (k) at (1.5,0)[right]{$k=q_1+q_2$};
    \diagram*{
    (p1) --[fermion, thick] (p1g) --[fermion, thick] (p1p);
    (p2) --[fermion, thick] (p2g) --[fermion, thick] (p2p);
    (p1g) -- [opacity=0,momentum={$q_1$}] (3g);
    (p2g) -- [opacity=0,momentum={[right]$q_2$}] (3g);
    };
\end{feynman}
    \node (3g) at (0.5,0){};
    \node (p2g) at (0,1){};
    \node (p1g) at (0,-1){};
    \node (k) at (1.5,0){};
    \draw[double, vector, line width = .75] (3g.center) -- (k.center);
    \draw[double, vector, line width = .75] (3g.center) -- (p1g.center);
    \draw[double, vector, line width = .75] (3g.center) -- (p2g.center);
    \node at (0.7,-0.3) {$I$};
\end{tikzpicture}
\caption{Five-point inelastic amplitude, the bottom (top) particle has mass $m_1$ ($m_2$) and spin vector $a_1$ ($a_2$). The interaction $I$ can be any of $I_1$, $G_3$ or $I_2$.}
\label{fig:5ptAmp}
\end{figure}

\subsection{The five-point spinning amplitude with cubic interactions}
We  now compute the five-point amplitude for the scattering of two objects with  masses $m_1$ and $m_2$ and spin vectors $\veca_1$ and $\veca_2$. We will obtain  the fully relativistic  amplitude, and then perform  PN expansion in order to extract the correction to the quadrupole term in the effective action \eqref{correctedL}. 

The relevant diagram is shown in Figure~\ref{fig:5ptAmp}, where we  use the three-graviton vertices \eqref{eq:VertexI1} and \eqref{eq:VertexI2}, 
the energy-momentum tensor \eqref{eq:Tmunu} and the graviton propagator in de Donder gauge.  
The momenta of the two internal gravitons are  spacelike, corresponding to an instantaneous interaction. The energy of the radiated (on-shell) graviton   is such that 
$k^0 \ll |\vec{q}_{1,2}\,|$, so that we can take the soft limit, $k\to 0$, and set \ $q_1\to -q$ and $q_2\to q$. In terms of length scales, this condition corresponds to demanding that the wavelength of the emitted radiation $\lambda \gg b$.
The emitted graviton thus enters the amplitude only through the  Riemann  tensor ${R}_{\alpha\beta\mu\nu}(k)$.
Doing so, we  obtain for the two interactions $I_1$ and $G_3$, 
\begin{align}
    \cM_{5,I_1} &= -48 i\left(\frac{\kappa}{2}\right)^4 \frac{q^\beta q^\delta}{q^2} {R}_{\alpha\beta\gamma\delta}(k)\bigg[2(p_1\Cdot J_1)p_2^\alpha J_2^\gamma + 2(p_2\Cdot J_2) p_1^\alpha J_1^\gamma \nn \\
    &\qquad - \Big((p_1\cdot p_2) J_1^\alpha J_2^\gamma + (p_1\Cdot J_2)J_1^\alpha p_2^\gamma + (p_2\Cdot J_1)p_1^\alpha J_2^\gamma + (J_1\Cdot J_2)p_1^\alpha p_2^\gamma\Big)\bigg], \label{eq:5ptAmpI1}\\
    \cM_{5,G_3} &= \cM_{5,I_1} - 24 i \left(\frac{\kappa}{2}\right)^4 \frac{q^\beta q^\delta}{q^2} {R}_{\alpha\beta\gamma\delta}(k)\Big[  (p_2\Cdot J_2) p_1^\alpha J_1^\gamma + (p_1\Cdot J_1)p_2^\alpha J_2^\gamma \Big]\ . \label{eq:5ptAmpG3}
\end{align}
Here we have defined the vectors
\begin{equation}
    J_i = p_i \cosh{(a_i\Cdot q)} + i e_i \frac{\sinh{(a_i\Cdot q)}}{a_i\Cdot q}\, , 
\end{equation}
and the Levi-Civita contractions
\begin{equation}\label{eqn:L-C Contractions}
    e_1^\mu \coloneq -\epsilon(\mu a_1p_1q) \, , \qquad e_2^\mu \coloneq \epsilon(\mu a_2 p_2 q)\ .
\end{equation}

\subsection{Spinning quadrupole moments}\label{sec:SpinningQuadrupoleMoments}
With the five-point amplitudes at hand, we can now extract the quadrupole moments from them,  by matching to the effective action  \eqref{correctedL}. In fact, the five-point amplitude $\cM_5$ contains more information than we need for our present purposes, as
\begin{equation}
    \cM_5 = \tilde{Q}^{ij}R^{0i0j}+\tilde{J}^{jk}\epsilon^{ik\ell}R^{0ij\ell} + \cdots\ .
\end{equation}
The first term corresponds to the quadrupole interaction, while the second corresponds to the current quadrupole, which is subleading in the multipolar expansion \cite{Maggiore:2007ulw,Goldberger:2009qd}. The dots represent higher order multipoles coming from subleading terms in the soft expansion of the amplitude. To match to the point-particle effective action \eqref{correctedL} we will need to Fourier transform and account for a non-relativistic normalisation factor, similar to the potential. For the case of extracting the quadrupole from the five-point amplitude, 
the appropriate integral is
\begin{equation}
    Q^{ij}(\vecr,\vecp) = -\int\!\frac{d^3\vecq}{(2\pi)^3} e^{-i \vecq\Cdot\vecr} \frac{i\tilde{Q}^{ij}(q)}{2E_1 E_2}\ ,
\end{equation}
with a different sign 
compared to \eqref{eqn:PotentialNormFactor} owing to the different signs of the second and third term  in the point-particle effective action \eqref{correctedL}. 
In evaluating the quadrupole part of the amplitude, we encounter within $J_i^0$ the $0^{\rm th}$ component of the Levi-Civita contractions of \eqref{eqn:L-C Contractions}. We can rewrite this by leveraging the fact that we are in the centre of mass frame, where $p_1 + p_2 = (E_1+E_2,\boldsymbol{0}) = (E_1+E_2)\hat{t}$, with $\hat{t}\coloneq (1, \boldsymbol{0})$. Thus, 
\begin{equation}
    e_1^0 = \hat{t}\Cdot e_1 = \frac{1}{E_1+E_2}(p_1 + p_2)\Cdot e_1 = \frac{p_2 \Cdot e_1}{E_1+E_2}\ ,
\end{equation}
since $p_1\Cdot e_1 = 0$. The Levi-Civita contractions defined earlier in \eqref{eqn:L-C Contractions} will now only appear in the following combinations: $p_2 \Cdot e_1$ and $p_1\Cdot e_2$. Reshuffling the indices we can  rewrite these as
\begin{align}
    p_2\Cdot e_1 &= -\epsilon(p_2 a_1 p_1 q) = -m_1 m_2 \epsilon(u_1 u_2 a_1 q) =: -m_1 m_2\ q\Cdot\epsilon_1\ , \\
    p_1\Cdot e_2 &= \epsilon(p_1 a_2 p_2 q) = -m_1 m_2\epsilon(u_1 u_2 a_2 q) =: -m_1 m_2\ q\Cdot\epsilon_2\ .
 \end{align}
In principle, the Fourier integrals can be done using methods identical to the potential integrals. However, we now have integrands with two spurious poles, corresponding to the $\sinh{(a_1\Cdot q)} \sinh{(a_2\Cdot q)}$ terms. This would require two Schwinger parameter integrals to calculate exactly. Since we are only interested in the leading term in  spin, we can expand the integrand in spin, greatly simplifying the calculations. Integrands with dot products of $q$ in the numerator, such as $a_i\Cdot q$ or $q\Cdot\epsilon_i$ can then be performed using the shifting trick in \eqref{eqn:DerivativeTrick}. These are, generically,
\begin{equation}
    \mathcal{Q}^{ij}_{n_1\cdots n_4} \coloneq \int\!\frac{\hat{d}^3\vecq}{(2\pi)^3} \frac{q^i q^j}{\vecq^2} (\veca_1\Cdot \vecq)^{n_1} (\veca_2\Cdot \vecq)^{n_2} (\veceps_1\Cdot \vecq)^{n_3} (\veceps_2\Cdot \vecq)^{n_4} e^{-i\vecq\Cdot\vecr}\ ,
\end{equation}
noting that at order $m$ in spin, we must have $\sum_i n_i = m$. Notice that when  expanding $\sinh{(a_i\Cdot q)}$, the spurious $a_i\Cdot q$ poles disappear. Now, we can express the integral above as derivatives of the shifted $\vecr$,
\begin{align}
    \mathcal{Q}^{ij}_{n_1\cdots n_4} &= \left.\frac{\partial^{n_1}}{\partial t_1^{n_1}}\cdots \frac{\partial^{n_4}}{\partial t_4^{n_4}}\int\!\frac{\hat{d}^3\vecq}{(2\pi)^3} \frac{q^i q^j}{\vecq^2} e^{-i\vecq\Cdot(\vecr+it_1\veca_1+\cdots+it_4 \veceps_2)}\right|_{t_i=0} \nn\\
    &= \frac{\partial^{n_1}}{\partial t_1^{n_1}}\cdots \frac{\partial^{n_4}}{\partial t_4^{n_4}} \mathcal{Q}^{ij}(\vecr+it_1\veca_1+\cdots+it_4 \veceps_2)|_{t_i=0}\ ,
\end{align}
where $\mathcal{Q}^{ij}$ is the quadrupole integral appearing in the scalar calculation,
\begin{equation}
    \mathcal{Q}^{ij}(\vecr) \coloneq \int\!\frac{\hat{d}^3\vecq}{(2\pi)^3} \frac{q^i q^j}{\vecq^2} e^{-i\vecq\Cdot\vecr} = -\frac{3}{4\pi \vecr^5}\Big(r^i r^j-\frac{1}{3}\vecr^2\delta^{ij}\Big)\ .
\end{equation}
The Fourier transforms can now be computed efficiently. Here we present the results up to and including spin squared terms. As in the calculation of  the potentials, we work in the aligned spin case and in the centre of mass frame. Expanding in PN we get
\begin{align}
\label{eqn:I1SpinningQuadrupole}
    Q^{ij}_{I_1} &=
    \left(\frac{\kappa }{2}\right)^4\frac{36 }{\pi  r^3}\Biggl[%
    \frac{\left(m_1+m_2\right)^2 }{m_1 m_2}\vecp^2 \hat{r}_{\langle i,j\rangle }+3 \left(a_1+a_2\right) \left(m_1+m_2\right) \sqrt{\vecp^2}\frac{ \hat{r}_{\langle i,j\rangle }}{r}\nn\\
    &+\left(\frac{5 a_1 a_2 m_1 m_2}{2 r^2}-\frac{\left(m_1^2-m_2^2\right){}^2 \vecp^4}{2 m_1^3 m_2^3}\right) \hat{r}_{\langle i,j\rangle }-\frac{a_1 a_2 m_1 m_2 \hat{z}_{\langle i,j\rangle }}{r^2}\Biggr]+\cdots\, , \\
    \label{eqn:G3SpinningQuadrupole}
    Q^{ij}_{G_3} &= -
    \left(\frac{\kappa }{2}\right)^4\frac{18 }{\pi  r^3}\Biggl[%
    m_1 m_2 \hat{r}_{\langle i,j\rangle }-\frac{2 \left(m_1+m_2\right)^2 \vecp^2 \hat{r}_{\langle i,j\rangle }}{m_1 m_2}\nn\\
    &-\frac{3 \sqrt{\vecp^2} \big(\left(4 a_1+3 a_2\right) m_1+\left(3 a_1+4 a_2\right) m_2\big) \hat{r}_{\langle i,j\rangle }}{2 r}+\Biggl(\frac{5 \left(a_1-a_2\right)^2 m_1 m_2}{2 r^2}\nn\\
    &+\frac{9 \left(m_1^2-m_2^2\right)^2 \vecp^4}{8 m_1^3 m_2^3}\Biggr) \hat{r}_{\langle i,j\rangle }-\frac{\left(a_1-a_2\right){}^2 m_1 m_2 \hat{z}_{\langle i,j\rangle }}{r^2}\Biggr]+\cdots\, . 
\end{align}
We have used the notation $X_{\langle i,j\rangle} {\coloneq} X_i X_j -\frac{1}{3}X^2 \delta_{ij}$ to denote the symmetric trace-free part, and we have aligned the spin of the two bodies along the  $\hat{z}$ axis.

A few comments are in order here. As already mentioned, we have organised the PN expansion in powers of   $\vecp^2 {\sim} 1/r {\sim} v^2$. The cubic corrections to  Einstein's quadrupole 
\begin{align}
\label{eqn:NewtonianQuadrupole}
Q^{ij}_N \coloneq \mu r^2 \hat{r}_{\langle i,j \rangle }
= \mu\Bigl(r^i r^j-\frac{1}{3}r^2\delta^{ij}\Bigr)
\, , 
\end{align}
with $\mu$ being the reduced mass, appear at 6PN and 5PN order for $I_1$ and $G_3$, respectively -- these are the same orders as in the corrections to the potential \eqref{eqn:I1SpinningPot} and \eqref{eqn:G3SpinningPot}, and for convenience we quote here terms up to 7PN.   
Furthermore, most of the $I_1$ and $G_3$ corrections are proportional  but there are additional contributions proportional to $\hat{z}_{\langle i,j \rangle }$. 
We also note that in the scalar case, the quadrupole correction   due to the $I_1$ interaction receives contributions  starting at $O(\vecp^2)$, while in the presence of spin there are corrections of $O(1)$ in the velocities but quadratic in the spins. 
Furthermore, the  quadrupole moment   receives $O(\sqrt{\vecp^2})$ corrections from  $I_1$ and $G_3$, for similar reasons as the potentials \eqref{eqn:I1SpinningPot} and~\eqref{eqn:G3SpinningPot}.

\subsection{On the relation between cubic and  tidal interactions}

In \cite{AccettulliHuber:2020dal} it was shown that an appropriate field redefinition of the metric 
\begin{align}
	g_{\alpha \beta} \rightarrow g_{\alpha \beta} 
    + \frac{3}{8} \kappa^2 \beta_2 \, g_{\alpha \beta} R^{\mu \nu}\,_{\rho \sigma} R^{\rho \sigma}\,_{\mu \nu}\ , 
\end{align}
maps the cubic interaction $G_3$ into  the first term in the action for the leading tidal deformations for scalar  fields 
introduced in \cite{Cheung:2020sdj}, 
\begin{align}
\label{tidal}
	S_{\rm tidal}= \int\!d^4x \sqrt{-g} \ 
    \frac{1}{4} R_{\mu \alpha \nu \beta} R^{\rho \alpha \sigma \beta}\sum_{i=1,2}\Big(\lambda_i \, \phi_i^2\delta^\mu_\rho \delta^\nu_\sigma + \frac{\eta_i}{m_i^4} \nabla^\mu \nabla^\nu \phi_i \nabla_\rho \nabla_\sigma \phi_i \,  \Big) 
	\ .
\end{align}
In the worldline formalism, these leading tidal deformations correspond to \cite{Cheung:2020sdj}
\begin{align}
\label{tidalWL}
	S_{\rm tidal}^{\rm WL}= \sum_{i=1}^2 \frac{1}{m_i}\int\!d\tau_i\, \Big[   2 \lambda_i\big( E^2_i  - B^2_i\big)  + \frac{\eta_i}{4}E_i^2 
    \Big] \ 
	\ , 
\end{align}
where the electric and magnetic components of the Weyl tensor are 
\begin{align}
E_{i,\mu\nu} &= v_i^\alpha v_i^\beta C_{\mu \alpha \nu \beta}\, ,  \\
B_{i,\mu\nu} &= \frac{1}{2}v_i^\alpha v_i^\beta {\epsilon_{\mu \alpha}}^{\rho \sigma} C_{\rho \sigma \nu \beta} \, . 
\end{align}
Hence, for non-spinning objects the $G_3$  deformation is   equivalent to the particular  tidal interaction  $E^2 {-} B^2 {\sim} R^2$. 
This has also been argued by \cite{Bern:2020uwk} on the basis that the Compton  amplitude produced by the $G_3$ interaction is a contact term%
\footnote{The term proportional to $q^2$ in the following equation gives rise to quantum corrections and can therefore be discarded. }
\cite{Brandhuber:2019qpg,Emond:2019crr} 
\begin{align}
\label{G3amplitude}
    \cM_{G_3}^{\rm scalar}(p, k_1^{++}, k_2^{++}) = -3 \, i \, \left( \frac{\kappa}{2}\right)^4  [1\, 2]^4 \left(m^2 + \frac{q^2}{2}\right) 
    \, , 
\end{align}
with $q=k_1 + k_2$. Because this term is local, it is already captured by the tidal operator of the form  $E^2 - B^2$. 

We can immediately argue that 
the same is true if the particles have spin: indeed, the Compton amplitude has now  the form \cite{Brandhuber:2024bnz}
\begin{align}
\label{G3amplitudewithspin}
    \cM_{G_3}(p, k_1^{++}, k_2^{++}) =  \cosh (a\Cdot q)   \cM_{G_3}^{\rm scalar}(p, k_1^{++}, k_2^{++}) 
    \, , 
\end{align}
which is also a local term. 
Furthermore, the  second term in \eqref{tidal} maps to $E_{\mu\nu} E^{\mu\nu}$ once we replace the derivatives $\nabla^\mu$ by the four-momentum $m v_i^\mu$ carried by the scalar, and is therefore an electric tidal deformation.
We are not aware of any field redefinition that could map this tidal deformation to a purely gravitation one.

However, we observed further similarities between tidal interactions and cubic deformations, which we wish to summarise briefly in the following.

Tidal effects for spinning particles were studied in detail in \cite{Aoude:2020ygw}. 
In that  paper a general formula for the impulse (momentum kick) was presented. Remarkably, the linear-in-spin terms (lines 2 and 3 of (5.4)) have exactly the same form as the linear-in-spin correction of the impulse arising from the other cubic deformation, $I_1$, which was presented in the first line of (6.65) of \cite{Brandhuber:2024bnz}. The first  ($S^0$) term in  (5.4) of \cite{Aoude:2020ygw} instead maps to our $G_3$ correction to the impulse, see (6.66) of \cite{Aoude:2020ygw}.

Finally, we point out an additional similarity (but not full equivalence) in the one-loop potential generated by the $I_1$ deformation and the leading magnetic tidal deformation. The latter can be found in  (A.3) and (A.7) of \cite{Bern:2020uwk},
with the electric and magnetic potentials being proportional to the coefficients
\begin{align}
\label{ele}
C_{E^2} = \frac{5}{2^3}G^2 m_1 m_2^3 \pi^2 \Big(\frac{11}{5} - 6\sigma^2 + 7 \sigma^4\Big)\, , 
\\
\label{magn}
C_{B^2} = \frac{5}{2^3}G^2 m_1 m_2^3 \pi^2 (\sigma^2-1)(1 + 7 \sigma^2)\ ,  
\end{align}
and note the $\sigma^2-1$ prefactor in  \eqref{magn}. Up to $O(\sigma^2)$, $C_{B^2}$ produces a potential that  is the same as the potential generated by the $I_1$ interaction, see (2.20) of \cite{Brandhuber:2019qpg}. This is because%
\footnote{Setting $\vecq^2=0$ to drop quantum corrections. }
\begin{align}
\sigma = \frac{\sqrt{{m_1}^2+\vecp^2}
   \sqrt{{m_2}^2+\vecp^2}+\vecp^2}{{m_1} {m_2}}\, , 
    \end{align}
so that to lowest order in $\vecp^2$
\begin{align}
  \sigma^2-1 \sim  \vecp^2 \frac{ (m_1+m_2)^2}{m_1^2 m_2^2}\, , 
\end{align}
  that is  $\vecp^2\sim  \sigma^2 -1$. However the subleading in $\vecp^2$ terms in the potential arising from the $I_1$ deformation do not capture the correction  proportional to $7\sigma^2 (\sigma^2-1)$ in \eqref{magn}.%
  \footnote{Also note that 
  $ C_{E^2} - C_{B^2}
 = C_{R^2} = 
  2\, G^2 m_1 m_2^3 \pi^2$, with the $\sigma$ dependence having cancelled from the result, giving the $G_3$ potential of (2.25) of \cite{Brandhuber:2019qpg}.}

\section{From amplitudes to waveforms}\label{sec:AmplitudesToWaveforms}
We have seen how cubic corrections modify various aspects of binaries dynamics. Through the potential, not only will the size of their circular orbits be modified, but so will the energy contained in their interaction. From the corrected quadrupoles, the power emission as given by  Einstein's quadrupole formula will also be modified. Combining these three things using energy conservation, we will be able to determine the corrections to the phasing of gravitational waves due to the $R^3$ interactions. 

Focusing on now on quasi-circular orbits, these corrections can be expanded in terms of the velocity, 
\begin{equation}
    v \coloneq (G M \Omega)^{1/3}\ .
\end{equation}
We will also find it useful to write expressions in terms of the following combinations of the masses $m_1$ and $m_2$:
\begin{equation}
    M \coloneq m_1 + m_2\ ,\qquad \mu \coloneq \frac{m_1 m_2}{M}\, ,\qquad \nu \coloneq \frac{m_1 m_2}{M^2}\ .
\end{equation}
Note that the symmetric mass ratio $0 {<} \nu {\leq} 1/4$, achieving its maximum when $m_1{=}m_2$.\\

\subsection{Corrections to circular orbits}\label{sec:CorrectionsRadius}
Due to the inclusion  of $R^3$ interactions, the size of the circular orbit at some angular velocity $\Omega$ will be modified compared to the Newtonian case, which is given by the well-known Kepler's law
\begin{equation}
    r_N^3 = \frac{G M}{\Omega^2}\, . 
\end{equation}
The size of the orbits will instead be given by some $r_{R^3} = r_N + \delta r$.
We can calculate this difference by examining the spinning $R^3$ potentials \eqref{eqn:I1SpinningPot} and \eqref{eqn:G3SpinningPot}. Generically, these corrections mean that the Hamiltonian takes the form
\begin{align}\label{eqn:Hamiltonian}
    H = \frac{\vecp^2}{2\mu}+V(r)+\epsilon\sum_{n=0}^N|\vecp|^n U_n(r)\ .
\end{align}
Here the small parameter $\epsilon$ will play the role of $\beta_{1,2}$, and $\vecp^2 = p_r^2 + p_\theta^2/r^2$ in plane polar coordinates. Since $H$ does not depend on $\theta$, we deduce that the angular momentum $p_\theta :=\ell$ is conserved. Imposing the first condition for a circular orbit, $\dot{r}=0$, leads to $p_r=0$. Looking at the equation of motion for $\dot{\theta}$,
\begin{align}\label{eqn:AngularEoM}
    \frac{\partial H}{\partial p_\theta} = \frac{p_\theta}{\mu r^2} + \epsilon \sum_{n=1}^N \frac{n p_\theta^{n-1}}{r^n} U_n(r) = \dot{\theta} := \Omega \ .
\end{align}
In general, this is a $n{-}1$ order polynomial equation for $p_\theta$. However, since we are only interested in the motion to linear order in $\epsilon$, we can solve the system approximately by writing
\begin{align}
    p_\theta = p_\theta^{(0)} + \epsilon\, p_\theta^{(1)} + O(\epsilon^2)\ .
\end{align}
This leads to the solution
\begin{align}
    p_\theta = \mu r^2\Omega - \epsilon\sum_{n=1}^N n \mu^n r^n \Omega^{n-1} U_n(r) + O(\epsilon^2)\ .
\end{align}
The final equation for the Hamiltonian system, for $\dot{p_r}$, will provide an equation for the radii of circular orbits. Explicitly,
\begin{align}
    -\frac{p_\theta^2}{\mu r^3} + V'(r) + \epsilon \sum_{n=0}^N \biggl( -\frac{n p_\theta^n}{r^{n+1}} U_n(r) + \frac{p_\theta^n}{r^n} U'_n(r) \biggr) = 0\ .
\end{align}
We can use the angular equation of motion \eqref{eqn:AngularEoM} to simplify the expression, leading~to
\begin{align}
    V'(r) - \frac{p_\theta}{r}\Omega + \epsilon\sum_{n=0}^N \frac{p_\theta^n}{r^n}U'_n(r) = 0 \ .
\end{align}
This can be solved for small perturbations about the Newtonian circular orbit $r_N$. Inserting the specific values for the potentials $U_n(r)$ from \eqref{eqn:I1SpinningPot} and \eqref{eqn:G3SpinningPot}, the corrections to the circular radius are, to linear order in the couplings $\beta_i$, 
\begin{align}
\label{modifr}
    \delta r = &\frac{576 \pi  v^8}{G^2 M^3} \biggl[\beta_1\left(\frac{2 v^2}{3}+\frac{6 v^3 \left(a_2 m_1+a_1 m_2\right)}{G M^2}-\frac{1}{6} (1-4 \nu) v^4\right)\nn\\
    &-\beta_2\biggl(1-\frac{1}{3} (1-2 \nu ) v^2+\frac{4 \left(a_1^2+a_2^2\right) v^4}{G^2 M^2}+\frac{1}{24} \left(8 \nu ^2-12 \nu +3\right) v^4\biggr)\biggr]+
\cdots    \, .
\end{align}

\subsection{Correction to the energy density}\label{sec:CorrectionsEnergy}
In unmodified gravity, the energy density contained in a binary system (the energy per unit mass $M$) is
\begin{equation}
    E = -\frac{G m_1 m_2}{2 M r} = -\frac{v^2\nu}{2}\ .
\end{equation}
Evaluating  the Hamiltonian \eqref{eqn:Hamiltonian}  in the presence of the cubic deformations and including the modifications to the radius of circular orbits found in \eqref{modifr}, the corrections to the energy density become
\begin{align}
\label{corrE1}
    \Delta E_{I_1}(v)&=\beta_1\frac{1056 \pi  \nu  \, v^{12}}{G^3 M^4}\biggl[v^2+\frac{72 v^3 (a_2 m_1+a_1 m_2)}{11 G M^2}-\frac{13}{22} (1-4 \nu) v^4\biggr]+ \cdots \\
    \label{corrE2}
    \Delta E_{G_3}(v)&= -\beta _2\frac{864 \pi \nu \,  v^{12}}{G^3 M^4}\bigg[ 1-\frac{11}{18} (1-2 \nu) v^2+\frac{13 (a_1^2+a_2^2) v^4}{3 G^2 M^2}
    \nn\\	&\hspace{80pt}
        +\frac{13}{72} (8 \nu ^2-12 \nu +3) v^4\bigg]  + \cdots 
\end{align}
We have expanded the energies up to 7PN for convenience, as done earlier for the potentials and quadrupole moments.

\subsection{Corrections to the power flux}\label{sec:CorrectionsPowerFlux}
The power emitted by the binary system is given by the famous Einstein quadrupole formula,%
\footnote{It would be interesting to see if it is possible to compute  the power from a continuation of the scattering  waveforms computed in   \cite{Brandhuber:2024qdn}, similarly to what was done in \cite{Falkowski:2024yuy}. }
\begin{equation}
\label{EQF}
    \mathcal{F} = \frac{G}{5}\langle \dddot{Q}^{ij} \dddot{Q}^{ij}\rangle\ .
\end{equation}
The quadrupole is averaged over one orbit of the binary, denoted by the angle brackets. Up to leading order in the $\beta_i$ and spin expansions, the quadrupole is,  limiting ourselves to linear order in the spins,  
\begin{align}
    Q = Q_{N} + Q_{R^3} + Q_{R^3, a^1} + O(\beta_i^2, a^2)\ ,
\end{align}
meaning that the full emitted power at this leading order is given by
\begin{equation}\label{eqn:FluxLinearInSpin}
    \mathcal{F}|_{R^3, a^1} = \frac{G}{5} \langle\dddot{Q}^{ij}_{N, a^0} \dddot{Q}^{ij}_{R^3, a^1}\rangle\ .
\end{equation}
Here, $Q_{R^3,a^1}$ are the quadrupoles given in \eqref{eqn:I1SpinningQuadrupole} and \eqref{eqn:G3SpinningQuadrupole}, and $Q_{N,a^0}$ is Einstein's  quadrupole given in \eqref{eqn:NewtonianQuadrupole}.

In principle, there should be a contribution to \eqref{EQF} arising from the overlap  of the leading spinning correction in general relativity  and the scalar $R^3$ corrections, however we will now show that this contribution vanishes. Working in the transverse-traceless ($TT$) gauge, the leading multipolar contribution to the field is given by
\begin{equation}\label{eqn:FieldQuad}
    h^{ij}_{TT}(u,r,\hat{n}) = \frac{\kappa}{16\pi r} \Lambda^{ij,kl} \ddot{Q}^{kl}\ ,
\end{equation}
leaving  implicit the instruction to evaluate it at retarded time $u$. Here $\Lambda^{ij,kl}(\hat{n}) \coloneq  P^{ik}P^{jl}-\frac{1}{2}P^{ij}P^{kl}$ is the transverse-traceless projector, with $P_{ij} {\coloneq} \delta_{ij}-\hat{n}_i\hat{n}_j$, and $\hat{n}$ denotes the direction of the radiation. On the other hand, the linear-in-spin correction to the field in general relativity takes the form \cite{Bautista:2021inx}
\begin{equation}\label{eqn:FieldGRSpin}
    h^{ij}_{TT,a^1}(u,r,\hat{n}) = \frac{\kappa}{16\pi r}\Lambda^{ij,kl}(\hat{n})\ 2\varepsilon^{abc}\hat{n}^{a}\Big[m_1 a_1^c(\delta^{bk}\dot{v}_1^{l}+\delta^{bl}\dot{v}_1^{k})+(1\leftrightarrow 2)\Big]\ ,
\end{equation}
corresponding to a current quadrupole contribution. Let us take a step back from the quadrupole formula and consider where it originates from. Expanding the Einstein Field Equations to second order in the perturbation $h_{\mu\nu}$, one can derive an effective 
energy-momentum tensor for the gravitational waves. The angular power distribution, associated to the $(0r)$ component of this energy momentum tensor can be expressed as \cite{Misner:1973prb}
\begin{equation}
    T^{\rm GW}_{0r} = \frac{d\mathcal{F}}{d\Omega} = \langle \dot{h}^{ij}_{TT}\dot{h}^{ij}_{TT} \rangle\ .
\end{equation}
To obtain the total power flux $\mathcal{F}$, this needs to be integrated over all angles (\textit{i.e}.~$\hat{n}$). However, notice that when performing the angular integrations on the contraction of \eqref{eqn:FieldQuad} and \eqref{eqn:FieldGRSpin}, one obtains a vanishing result:
\begin{equation}
    \mathcal{F}\sim\int\! d\Omega\ \Lambda^{ij,kl}\Lambda^{ij,mn}\hat{n}^a = \int\! d\Omega\ \Lambda^{kl,mn}\hat{n}^a = \int\! d\Omega\ \hat{n}^k \hat{n}^l \hat{n}^m \hat{n}^n \hat{n}^a + \cdots = 0\ .
\end{equation}
This is because each of the integrals is fully symmetric, and thus can only depend on $\delta^{ij}$, but they all have an odd number of indices. This is a specific manifestation of a more general feature of the multipolar expansion \cite{Maggiore:2007ulw}. The field $h_{\mu\nu}$ can be expanded in terms of two families of symmetric trace-free (STF) tensors, often called the \textit{mass multipoles} and the \textit{current multipoles}, given in terms of integrals of the mass-energy distribution $T_{\mu\nu}$. It can be shown by symmetry that, for the power emitted, only time averages of the  multipoles with themselves contribute. This means that there are never any cross terms between mass and current quadrupoles, or between quadrupoles corresponding to different values of $\ell$.  Hence, we will only consider the linear-in-spin corrections from here on. In the ensuing calculations, this will translate to expanding to 6.5PN order, dropping the 7PN contributions where quadratic-in-spin terms begin. 

Returning to the expression for the flux in \eqref{eqn:FluxLinearInSpin}, we can perform the time averaging between the spinning $R^3$ quadrupoles of \eqref{eqn:I1SpinningQuadrupole} and \eqref{eqn:G3SpinningQuadrupole} and the Newtonian quadrupole \eqref{eqn:NewtonianQuadrupole}. We do this by parameterising the circular orbit with $\vecr = r(\cos{\Omega t},\sin{\Omega t},0)$ and averaging over one circular orbit,
\begin{equation}
    \langle \dddot{Q}^{ij}_{N, a^0} \dddot{Q}^{ij}_{R^3, a^1}\rangle \coloneq \frac{1}{2\pi/\Omega}\int_{0}^{2\pi/\Omega}
    \!dt\ \dddot{Q}^{ij}_{N, a^0} \dddot{Q}^{ij}_{R^3, a^1}\ .
\end{equation}
Doing so, one gets 
\begin{align}
\label{corrFlux}
	\mathcal F(v) &= \frac{32 \nu ^2 v^{10}}{5 G} + \beta_1\, \mathcal{F}_{I_1}(v) + \beta_2\,\mathcal{F}_{G_3}(v) \, , 
    \end{align}
    with 
    \begin{align}
	\mathcal{F}_{I_1}(v)&=\frac{196608 \pi  \nu ^2 v^{20}}{5 G^4 M^4}\biggl[v^2+\frac{9 v^3 ((a_1+2 a_2) m_1+(2 a_1+a_2) m_2)}{4 G M^2}]\biggr] \, , \\
\mathcal{F}_{G_3}(v)&= -\frac{147456 \pi  \nu ^2 v^{20}}{5 G^4 M^4} \biggl[1-\frac{1}{6} (7-2 \nu) v^2
	-\frac{3 v^3 ((4 a_1+3 a_2) m_1+(3 a_1+4 a_2) m_2)}{4 G M^2}\biggr]\, . 
\end{align}

\subsection{Corrections to the gravitational waveforms}\label{sec:CorrectionsWaveform}

Finally, in this section we focus on the 
the quasi-circular, adiabatic inspiraling of a binary black hole system, and compute the gravitational waveforms in the  PN approximation.
In frequency space, the waveform of a binary with orbital phase $\phi(t)$  is given by%
\footnote{Recall  that the gravitational-wave phase is twice the orbital phase $\phi$.}
\begin{equation}
    \tilde{h}(f) = \int\!dt\ A(t) e^{-{2}i\phi(t)} e^{2\pi ift}\ .
\end{equation}
We can evaluate this by making use of the saddle point approximation, when the phase 
\begin{align}
    \psi(t)\coloneq 2\pi ft - {2}\phi(t)\, , 
\end{align} 
in the integral is stationary,
\begin{equation}
\label{stationary}
    \dot{\psi}(t_f)=0\ .
\end{equation}
This leads  to the stationary phase approximation (SPA) result \cite{Damour:1997ub,Damour:2000zb,Buonanno:2009zt,Sennett:2019bpc}
\begin{equation}
    \tilde{h}_{\rm SPA}(f) = \frac{A(t_f)}{\sqrt{\dot{F}(t_f)}}\exp \Big[ {i\left(\psi_{\rm SPA}(t_f)-\frac{\pi}{4}\right)}\Big]\ ,
\end{equation}
where $F(t) {\coloneq} \dot{\phi}(t)/\pi$ is the instantaneous frequency of the emitted gravitational wave.
During the inspiral phase, where the motion is quasi-circular and adiabatic, good approximations to $t_f$ and $\psi_{\rm SPA}(t_f)$ are then given by%
\footnote{The following two equations arise from solving the differential equations
$\dot{\phi} = v^3 / (GM) $ and $\dot{v} = - \mathcal{F}(v) / (M E^\prime (v))$, valid for circular orbits, with $v=r\dot{\phi} $.  The first of these relations is  the  Euler-Lagrange equation for $r$ specified to  the case of circular orbits. The second is obtained simply from the definition of $\mathcal{F}$ and by writing $\dot{E} = E^{\prime}(v) \dot{v}$.
}
\begin{align}
    t_f &= t_{\rm ref} + M\int_{v_f}^{v_{\rm ref}}\! dv\ \frac{E'(v)}{\mathcal F(v)}\ , \\
    \label{psiSPA}
    \psi_{\rm SPA}(t_f) &= 2\pi f t_{\rm ref} - {2}\phi_{\rm ref} + \frac{2}{G}\int_{v_f}^{v_{\rm ref}}\! dv\ (v_f^3-v^3)\frac{E'(v)}{\mathcal F(v)}\ .
\end{align}
Here, $t_{\rm ref}$ and $\phi_{\rm ref}$ are integration constants, $v_{\rm ref}{\coloneq}v(t_{\rm ref})$, and $v_f{\coloneq} v(t_f) {=} (\pi GMf)^{1/3}$; and finally 
$\mathcal{F} {\coloneq} - M\,  dE/dt$, where $E$ is the total energy per unit mass.%
\footnote{The fact that  $
v(t_f) = (\pi GMf)^{\frac{1}{3}}$ follows from \eqref{stationary} and equation  $\dot{\phi}=v^3/ (GM)$ given earlier.}
Using \eqref{psiSPA} we
can now  obtain  the correction to  $\psi_{\rm SPA}(t_f)$ due to the cubic perturbations $I_1$ and $G_3$. To do so we perform a PN expansion of  the ratio  ${E'(v)}/ {\mathcal F(v)}$, using the results obtained earlier for the energy corrections \eqref{corrE1} and \eqref{corrE2} and the flux \eqref{corrFlux}. 
We then have 
\begin{align}
    \label{eqn:WaveformCorrectionsComplete}
    \psi_{\rm SPA}(t_f) &= 2\pi f t_{\rm ref} - {2}\phi_{\rm ref} +\frac{3}{128 v_f^5 \nu} + \beta_1\,\psi^{I_1}_{\rm SPA}(t_f)+ \beta_2\,\psi^{G_3}_{\rm SPA}(t_f) \, , 
\end{align}
    where for our cubic corrections we have
\begin{align}
\label{eqn:WaveformCorrectionI1}
    \psi^{I_1}_{\rm SPA}(t_f)&=-\frac{4905 \pi  v_f^5}{7 G^3 \nu  M^4}\biggl[v_f^2+\frac{126 \left(\left(2 a_1+19 a_2\right) m_1+\left(19 a_1+2 a_2\right) m_2\right) v_f^3}{545 G M^2}\biggr]\, , \\
    \label{eqn:WaveformCorrectionG3}
    \psi^{G_3}_{\rm SPA}(t_f)&=\frac{1404 \pi  v_f^5}{G^3 \nu  M^4}\biggl[1+\frac{5 (170 \nu -133) v_f^2}{2184}-\frac{3 \left(\left(4 a_1+3 a_2\right) m_1+\left(3 a_1+4 a_2\right) m_2\right) v_f^3}{52 G M^2}\biggr]\ .
\end{align}

We have ignored all terms depending on $v_{\rm ref}$ as these are reabsorbed into $t_{\rm ref}$ and~$\phi_{\rm ref}$. These results agree with those of \cite{AccettulliHuber:2020dal} in the case of spinless bodies ($a_1 = a_2 =0$).

\section*{Acknowledgements}
We would like to thank  
Gang Chen, Panagiotis Marinellis, Gustav Mogull, Nathan Moynihan and Paolo Pichini for  useful discussions. This work was supported by the Science and Technology Facilities Council (STFC) Consolidated Grant ST/X00063X/1 \textit{``Amplitudes, Strings  \& Duality''}.
The work of  PVM is supported by a STFC quota studentship.
The work of GRB is supported by the U.K. Royal Society through Grant URF{\textbackslash}R1 {\textbackslash}20109.
GT is also supported by a Leverhulme research fellowship RF-2023-279$\backslash 9$.
No new data were generated or analysed during this study.

\newpage

\appendix

\section{Fourier integrals}
\label{sec:FourierIntegrals}
The expressions for the integrals \eqref{eq:FourierIntegral1} and \eqref{eq:FourierIntegral2} are
\begin{align}
    \int\! \frac{d^3\vecq}{(2\pi)^3} e^{-i\vecq\Cdot\vecr} |\vecq|^3 &= \frac{12}{\pi^2} \frac{1}{|\vecr|^6}\ ,\\
    \int\! \frac{d^3\vecq}{(2\pi)^3} e^{-i\vecq\Cdot\vecr} |\vecq|^3\frac{\sinh{(\veca_2\Cdot \vecq)}}{\veca_2\Cdot\vecq} &= \frac{6i \veca_2\Cdot\vecr_+ \left(2N-3\veca_2^2 \vecr_+^2\right)}{8\pi^2 \vecr_+^4 N^2} + c.c.\nn\\
    &+ \frac{9\veca_2^4}{8\pi^2(-N)^{5/2}}\log{\left(\frac{\vecr^2+\veca_2^2+2\sqrt{-N}}{\vecr^2+\veca_2^2-2\sqrt{-N}}\right)}\ ,
\end{align}
where $\vecr_{\pm} = \vecr \pm i\veca_2$ and $N = (\veca_2\Cdot\vecr)^2-\veca_2^2\vecr^2<0$ by the Cauchy-Schwarz inequality.

\section{The $I_2$ vertex}
\label{sec:I2vertex}
 Below we give the expression for the $I_2$ vertex with two off-shell legs: 
\begin{equation}
\begin{split}
\label{eq:VertexI2} 
\begin{tikzpicture}[line width=1.5 pt,scale=.5,baseline={([yshift=-0.4ex]current bounding box.center)}]
	\draw[double, vector, line width = .75] (0:0) -- (0:2);
	\draw[double, vector, line width = .75] (0:0) -- (120:2);
	\draw[double, vector, line width = .75] (0:0) -- (240:2);
        \node at (60:1.2) {$I_2$};
        \node at (120:2.6) {$1,\alpha\beta$};
        \node at (240:2.6) {$2,\gamma\delta$};
\end{tikzpicture}
= 
6 i \left(\frac{\kappa}{2}\right)^2  &\Big(\eta ^{\alpha  \nu}\eta^{\gamma  \mu } k_1\cdot k_2+\eta ^{\alpha  \nu } k_1^{\gamma }
   k_1^{\mu }-k_1^{\nu } \big(\eta ^{\alpha  \gamma } k_1^{\mu }+\eta ^{\gamma  \mu } k_2^{\alpha }\big)\Big) \\[-4ex]
  \times &\Big(\eta ^{\beta  \sigma
   }\eta^{\delta  \rho } k_1\cdot k_2+\eta ^{\beta  \sigma } k_1^{\delta } k_1^{\rho }-k_1^{\sigma } \big(\eta ^{\beta  \delta } k_1^{\rho }+\eta
   ^{\delta  \rho } k_2^{\beta }\big)\Big)
   R_{\mu \nu \rho \sigma }\,.
   \end{split}
\end{equation}

\newpage

\bibliographystyle{JHEP}
\bibliography{ScatEq.bib}

\end{document}